\documentclass[onecolumn,amsmath,amssymb,prfluids,reprint]{revtex4-2}

\usepackage{subfigure}
\usepackage{graphicx}
\usepackage{dcolumn}
\usepackage{bm}
\usepackage{color}
\usepackage{multirow}

\renewcommand{\chi}{\mathcal{X}}

\newcommand{\vu}{{\bm u}}

\newcommand{\vtx}{{\omega}}
\newcommand{\vx}{{\bm x}}

\newcommand{\curl}{\nabla \times}

\newcommand{\grad}{\nabla}
\newcommand{\Laplace}{\Delta}

\def\imagetop#1{\vtop{\null\hbox{#1}}}

\begin{document}
\newcommand{\blue}[1]{\textcolor{blue}{#1}}


\title[]{The influence of the vorticity-scalar correlation on mixing}

\author{Xi--Yuan Yin$^1$, Wesley Agoua$^1$, Tong Wu$^{1,2}$, Wouter J.T. Bos$^1$}

\affiliation{$^1$ LMFA-CNRS, Ecole Centrale de Lyon, Universit\'e de Lyon,
  Ecully, France\\
$^2$ Theoretical Physics I, University of Bayreuth, Bayreuth, Germany}

\date{\today}

\begin{abstract}
We investigate the role of the correlation between a scalar quantity and the vorticity in two-dimensional mixing at infinite Péclet number. We assess, using a diffusivity independent mixing-norm, the dynamics of both Galerkin-truncated ensembles and freely evolving two-dimensional scalar mixing. Both statistical mechanics and numerical experiments show how the mixing-rate is attenuated when vorticity and scalar are initially correlated. Since the vorticity is shown to be a poorly mixing scalar, the results suggest that, in general, mixing can be enhanced by minimizing the correlation between vorticity and passive scalar.   
\end{abstract}

\maketitle

\paragraph{Introduction.}

The present investigation considers the mixing (or stirring) of a passive scalar.
The scalar can represent temperature fluctuations, pollutants or any other passively advected quantity.
The propension of a given flow to mix, is determined by the dimensionality of space, the properties of the velocity field and of the scalar, boundary conditions, the type of stirring, and initial conditions. We discuss these different features now, for the special case of two dimensions, considering an incompressible fluid, advecting a passive scalar. 

The influence of boundary conditions is clearly very important, as was shown by Raynal and Gence \cite{raynal1997energy} comparing chaotic mixing in periodic and unbounded domains. Chaotic mixing in wall bounded domains was considered by Gouillart et al. \cite{Gouillart2007,gouillart2008slow}, where the presence of unmixed fluid near solid walls was shown to slow-down the mixing. In turbulent flows, the generation of enstrophy near the walls was shown, on the contrary, to enhance mixing \cite{kadoch2020efficiency}. In the present case we will focus on the simplest academic case of periodic boundary condition, avoiding thereby the issues related to the influence of solid walls. 

The stirring protocol is another important factor that we will omit from our analysis. Indeed, to obtain efficient mixing, the stirring protocol can be optimized under constraints \cite{foures2014optimal, eggl2020mixing}.  We will let the flow freely evolve from initial conditions. The precise type of the initial conditions, in particular the typical size of the initial scalar compared to the typical velocity length scales, or domain-size can importantly influence the mixing \cite{schekochihinDiffusionPassiveScalar2004,haynes2005controls}. Besides, the study of strange eigenmodes reveal the existence of scalar fields of slowest decay which attract the long-time evolution of general initial scalar distributions \cite{liuStrangeEigenmodesDecay2004, sukhatme2002decay}, though the rate of decay of scalar variance is still sensitive to initial conditions.  Again, to keep the description as general as possible, we will consider random initial velocity and scalar fields. These fields are constituted by random Fourier modes with a prescribed energy spectrum for the velocity field, and analogously, a prescribed variance spectrum for the scalar quantity.

We have at this point simplified and specified the flow, the boundary conditions, omitted stirring and prescribed random initial conditions. The problem is then determined, except for  one point: to what extent the initial velocity and scalar fields are correlated? The influence of this initial correlation, more specifically the vorticity-scalar correlation is the main subject of this study. We will predict theoretically and assess by numerical experiments how the initial vorticity-scalar correlation affects the mixing efficiency. This question has received little attention \cite{holloway1984stirring,Lesieur1985} and only a very partial understanding of the influence of the scalar-vorticity correlation on mixing exists. In particular, the question whether such an initial correlation is beneficial or, on the contrary, detrimental for mixing has not been answered.

\paragraph{Governing equations and constants of motion.}

We consider the advection of a passive scalar field $\phi(\vx, t)$ in a 2D periodic box $\Omega$. The scalar is transported by a two-dimensional incompressible and inviscid flow whose velocity field $\vu(\vx, t)$ is described by the Euler equations. The curl of the velocity, or vorticity, $\vtx(\bm x,t)=\grad \times \vu(\vx, t)$ evolves according to 
\begin{subequations} \label{eqns:Euler}
\begin{align}
(\partial_t + \vu \cdot \grad) \vtx = 0\\
\vu = - \curl \Laplace^{-1} \vtx
\end{align}
\end{subequations}
where the velocity can be recovered from the vorticity by the Biot-Savart  relation Eq.~(1b). The initial condition is $\vtx (\vx, 0) = \vtx_0 (\vx)$. 

The linear advection equation under the velocity field $\vu$ then describes the evolution of the scalar $\phi$:
\begin{equation}\label{eqns:scalar}
 (\partial_t + \vu \cdot \grad) \phi = 0,
\end{equation}
with $\phi(\vx, 0) = \phi_0(\vx)$.
As we wrote above, we consider a general case where the initial velocity and scalar field $\vtx_0 (\vx),~\phi_0(\vx)$  are random. The investigation of more specific initial conditions is left for further research.

The quadratic invariants, or constants of motion, of a two-dimensional Euler velocity field  are  
\begin{eqnarray}\label{eq:invariants}
E=\int |\bm u|^2 d \vx, ~~~W=\int \vtx^2 d \vx.
\end{eqnarray}
The energy $E$ and enstrophy $W$ characterize the velocity field. One can define a typical wavenumber
\begin{equation}\label{eq:ku}
k_u=\sqrt{W/E}.
\end{equation}
This wavenumber characterizes the typical lengthscale of the velocity field. In the absence of sources and sinks, it remains constant. The scalar field is also characterized by two invariants,
\begin{eqnarray}\label{eq:invariants}
S=\int \phi^2 d \vx, ~~~Q=\int \vtx \phi d \vx.
\end{eqnarray}
Whereas the variance $S$ is determined by the scalar field only, the last quantity $Q$ measures the correlation between the scalar and the vorticity field.

Two-dimensional turbulence contains an infinite number of additional invariants, Casimirs of the form $\int f(\vtx) d \vx$. Equivalently for the scalar, additional invariants can be defined. The investigation of the influence of these invariants on mixing will be left for further research.  The choice to retain only the 4 quantities $E,W,S,Q$ is theoretically supported by the fact that in the Galerkin truncated system that we will consider first, they are the only surviving constants of motion, conserved, triad-by-triad by the nonlinear and advection terms \cite{fyfe1977dissipative,kraichnan1980two}. Clearly since we ignore all the Casimirs, our investigation can be refined in future studies.

The quantity $Q$ is, like the other invariants, conserved in the absence of sources or sinks, so that its initial correlation persists throughout the evolution of the system. We can further define the relative correlation,
\begin{equation}
h=Q/\sqrt{SW},
\end{equation}
a quantity which is constrained by the bounds $0\leq h\leq 1$.  Our goal is to measure and understand the influence of $h$ on mixing. 

\paragraph{Mixing norm.}

The eventual faith of a scalar in a realistic flow is determined by the diffusion. However, the mixing is, at sufficiently small values of the diffusivity, piloted by the stretching and folding of the large-scale scalar structures. 
Therefore we want to measure the mixing independently from the values of diffusion (and viscosity). We use a mixing norm previously introduced in the study of optimal mixing \cite{mathew2005multiscale} (see also \cite{thiffeault2012using}).
 This measure is a function of the inverse gradient $|\nabla^{-1}\phi|$; its value is representative of the importance of the large scales of $\phi$. We introduce a characteristic wavenumber,
\begin{equation}
k_\phi=\sqrt{\frac{\int \phi^2 d \vx}{\int |\nabla^{-1}\phi |^2 d \vx}}.
\end{equation}
An increase of this characteristic wavenumber indicates that the scalar is transported towards smaller scales. Typically this implies that the flow is mixing, since the eventual scalar dissipation in a realistic flow is taking place at the small scales (or large wavenumbers). We also note that in a periodic domain, for scalar $\phi$, one can check by Parseval's identity that $\int | \nabla^{-1} \phi |^2 d \vx = \sum_{\bm{k}} k^{-2} |\hat{\phi}(\bm{k}) |^2 = \int | -\nabla\times \Delta^{-1} \phi |^2 d \vx$ so that the typical wave number $k_u$ is the ratio of the $L^2$ to $H^{-1}$ seminorms for $\omega$, and $k_\phi$ is the same ratio for $\phi$.

Since the Euler-equations conserve both energy and enstrophy, $k_u$ (defined in Eq.~\eqref{eq:ku}) does not evolve. However $k_\phi$ can evolve and this evolution will characterize the mixing rate. In the following, we will investigate the evolution of $k_\phi$ for given values of $h$ and $k_u$. We will start by deriving a theoretical prediction.

\paragraph{Correlated and uncorrelated scalars.}

The passive scalar and the vorticity share the same evolution equation \eqref{eqns:scalar},\eqref{eqns:Euler}. The difference is that the vorticity is related to the velocity by Eq.~(\ref{eqns:Euler}b).
Since the evolution-equation is identical, we can decompose the scalar into a correlated part which is proportional to the vorticity and an uncorrelated part $\phi'$, 
\begin{equation}\label{eq:phicorr}
\phi(\bm x,t)= (h\sqrt{S/W})\vtx(\bm x,t) +\phi'(\bm x,t),
\end{equation}
with the following relations 
\begin{equation}
\int \phi' \vtx d \vx=0 \textrm{~~~and~~~}
S'\equiv\int \phi'^2 d \vx=S(1-h^2).
\end{equation}
The quantities $S$ and $S'$ are independently conserved. The decomposition of the scalar \eqref{eq:phicorr} allows to simplify the problem and to formulate predictions for the statistical mechanics of the scalar using known results only. Indeed, the evolution of $\phi'$ to a statistical equilibrium state is governed by one invariant, $S'$. The evolution of the correlated part will be fully determined by the evolution of $\vtx$, since $h$ is constant.

\begin{figure}
\centering
\begin{tabular}[t]{c c}
     \imagetop{(a)} &  \imagetop{\includegraphics[width=0.3\textwidth]{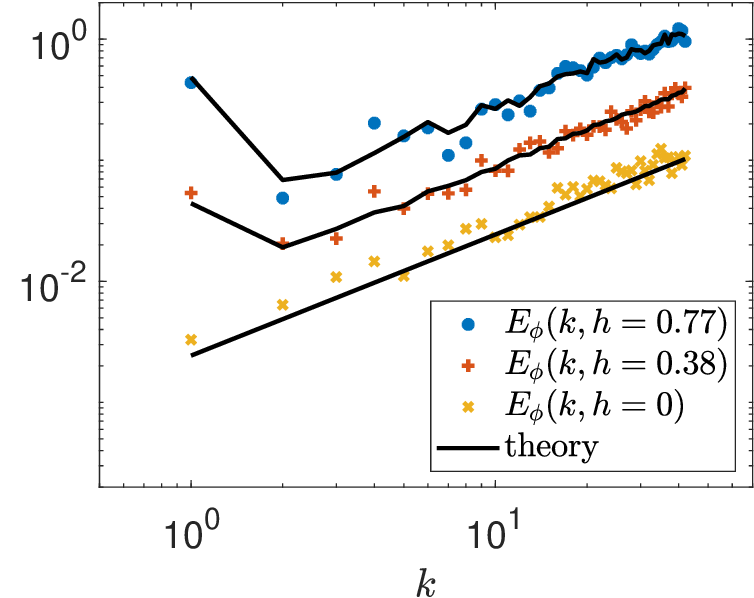}} \\
     \imagetop{(b)} & \imagetop{\includegraphics[width=0.3\textwidth]{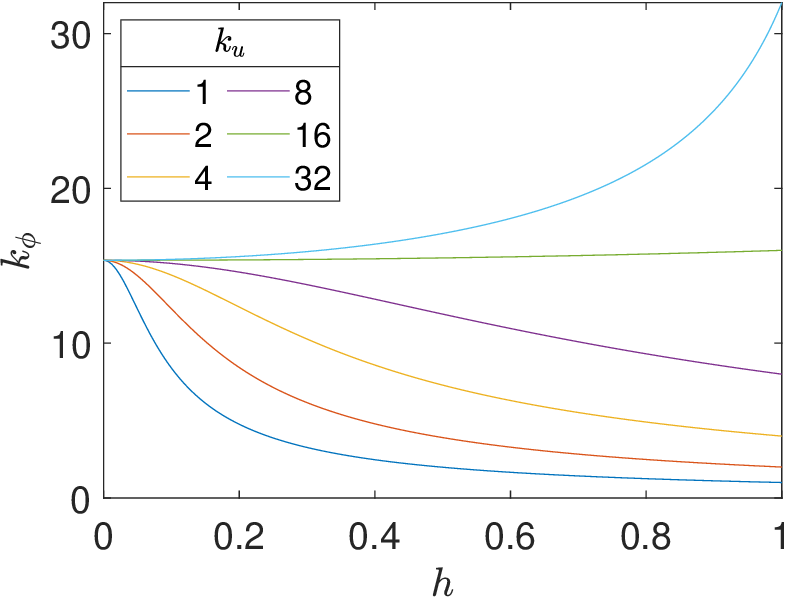}}
\end{tabular}
\caption{\small Results and predictions for equilibrium ensembles of the Galerkin-truncated system. The wavenumber range is confined to $k\in[1,42]$.
(a)DNS results for the Equilibrium spectra of the scalar variance $E_\phi(k)$. Theory predicts a spectrum of $h^2(S/W) k^2 E(k) + c_\phi k$ from eq. \eqref{eq:ESM}. For clarity of the image, the $h=0.38$ and $h=0.77$ data are shifted up by a half and full decade respectively. (b) 
Influence of the relative correlation $h$ on the mixing scale $k_\phi$ predicted from eq. \eqref{eq:kphiESM}. Results are shown for different values of the characteristic velocity scale $k_u$.}
\label{fig:ESM}
\end{figure}

\paragraph{Theoretical predictions for the Galerkin truncated system.}

We consider that the velocity and scalar excitation are projected on a finite number of Fourier modes in two-dimensional Fourier-space. The smallest and largest wavenumber are $k_0,k_1$ respectively. The evolution-equations are also restricted to keep the excitation confined on these modes. The scalar variance is then obtained by
\begin{equation}
\int_{k_0}^{k_1}E_\phi(k)dk=S.
\end{equation}
And a similar expression relates $E_{\phi'}(k)$ and $S'$.
In such a set-up, the equilibrium solution of the uncorrelated passive scalar is given by the equipartition solution
\begin{equation}
E_{\phi'}(k,t=\infty)=c_\phi k,
\end{equation}
where $c_\phi=2S'/(k_1^2-k_0^2)$ (see for instance \cite{Leith3}). This simple expression describes the equilibrium spectrum associated with $\phi'$. 
The correlated part is fully determined by the vorticity field and its dynamics is determined by both the kinetic energy and the enstrophy. Indeed, the energy spectrum determines both quantities according to 
\begin{equation}
\int_{k_0}^{k_1}E(k)dk=E,~~~\int_{k_0}^{k_1}k^2 E(k)dk=W.
\end{equation}
For the scalar spectrum we have therefore
\begin{equation}\label{eq:ESM}
E_{\phi}(k,t)=E_{\phi'}(k,t)+h^2 \frac{S}{W} k^2 E(k,t).
\end{equation}
To verify this relation, we carry out numerical simulations of the system. Standard Cartesian pseudospectral solvers are the ideal tools to verify the present predictions of equilibrium statistical mechanics, since these methods are based on a Galerkin truncation \cite{Orszag1969} and since they are known to correctly conserve the invariants of the system. We use the two-dimensional code Ghost \cite{mininni2011hybrid}, based on a vorticity-streamfunction formulation and second order time-integration. We start from a Gaussian-shaped initial spectrum $E(k,0)\sim  \exp(-b(k-k_L)^2)$, with random Fourier-phases and $k_L=4$ and $b=0.125$, setting the total energy at 1 and enstrophy at approximately 28, i.e. $k_u \approx 5.3$.

We plot in Fig.~\ref{fig:ESM}(a) the spectra $E_\phi(k,t=t_\infty)$ and the estimate Eq.~\eqref{eq:ESM}, with $E_{\phi'}(k)=c_\phi k$. For each value of $h(t=0)$ a different statistical realizations of the initial conditions for $E(k,0)$ is used. The time $t_\infty$ is determined by running the simulations for $t \approx 8000 t_*$, with $t_*=W^{-1/2}$,
a time sufficient for the statistics to converge to a statistical equilibrium. It is observed that the predicted spectrum is well approached by the theoretical estimate. We can then use this theoretical estimate of the scalar spectrum to compute analytically the mixing scale $k_\phi$.

Using Parseval's relation between the mixing norm and the scalar spectrum $ \int |\nabla^{-1}\phi |^2d \vx=\int_{k_0}^{k_1} k^{-2} E_\phi(k) dk$, we find after some elementary manipulation for the mixing scale, 
\begin{equation}\label{eq:kphiESM}
k_\phi=\left[h^2 k_u^{-2}+(1-h^2)k_{\phi'}^{-2}\right]^{-1/2}.
\end{equation}
where $k_{\phi'}=\left( (k_1^2-k_0^2)/(2\ln(k_1/k_0)) \right)^{1/2}$. The two limits of these expression are, for $k_\phi(h=1)=k_u$, where the mixing scale is fully determined by the typical velocity length-scale. And for the uncorrelated case we have $k_\phi(h=0)=k_{\phi'}$. 

Expression \eqref{eq:kphiESM}  is plotted for fixed resolution (or $k_{\phi'}$) in Fig.~\ref{fig:ESM}(b). We see there that the influence of finite $h$ is for most cases deleterious, i.e, $k_\phi$ is a decreasing function of $h$. Only for large initial $k_u$, this effect is not observed. Nevertheless, this latter case corresponds to initial conditions where the initial scalar field is concentrated at scales much larger than the excited velocity scales. In realistic flows this will lead to very slow mixing. The reason that statistical mechanics predict enhanced mixing for this case is that only final states are predicted, irrespective of the time it takes to reach such an equilibrium state. This constitutes the limit of equilibrium statistical mechanics to predict realistic flows and we will therefore in the following use numerical simulations to understand the dynamics.

\paragraph{Simulations of scalar mixing by the Euler-equations.}

\begin{figure*}
  \centering
  \begin{minipage}{0.68\textwidth}
  \subfigure[]{\includegraphics[width=0.45\textwidth]{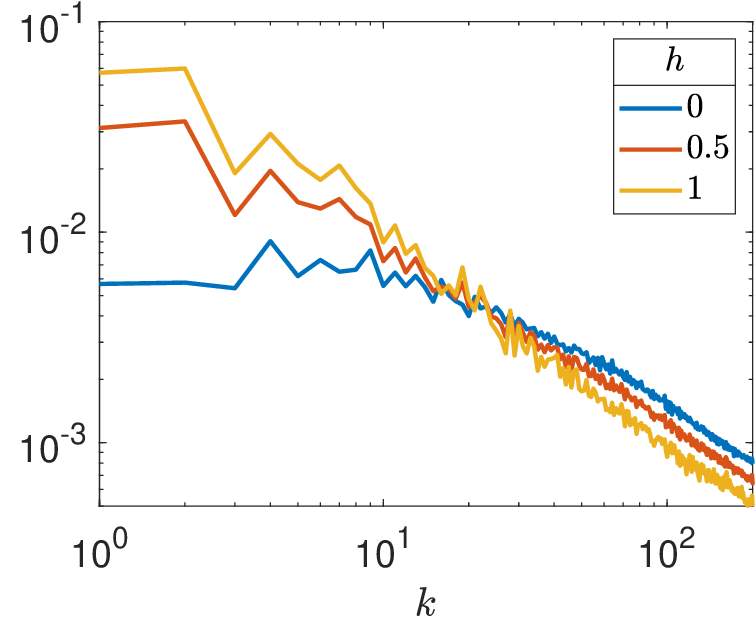}}  
  \subfigure[]{\includegraphics[width=0.52\textwidth]{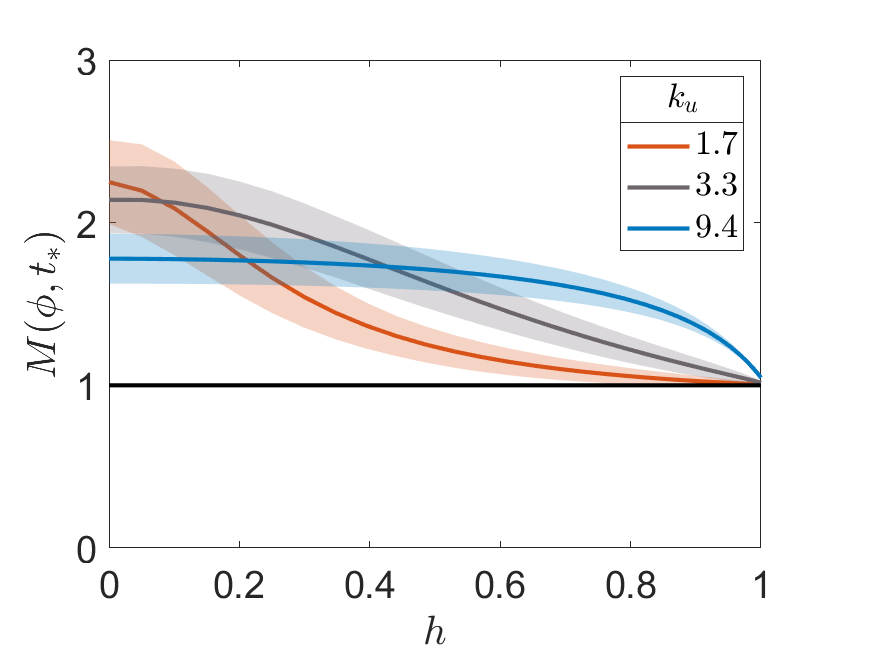}}
  \end{minipage}
  \begin{minipage}{0.22\textwidth}
  \subfigure[]{
    	\includegraphics[width=0.5\linewidth]{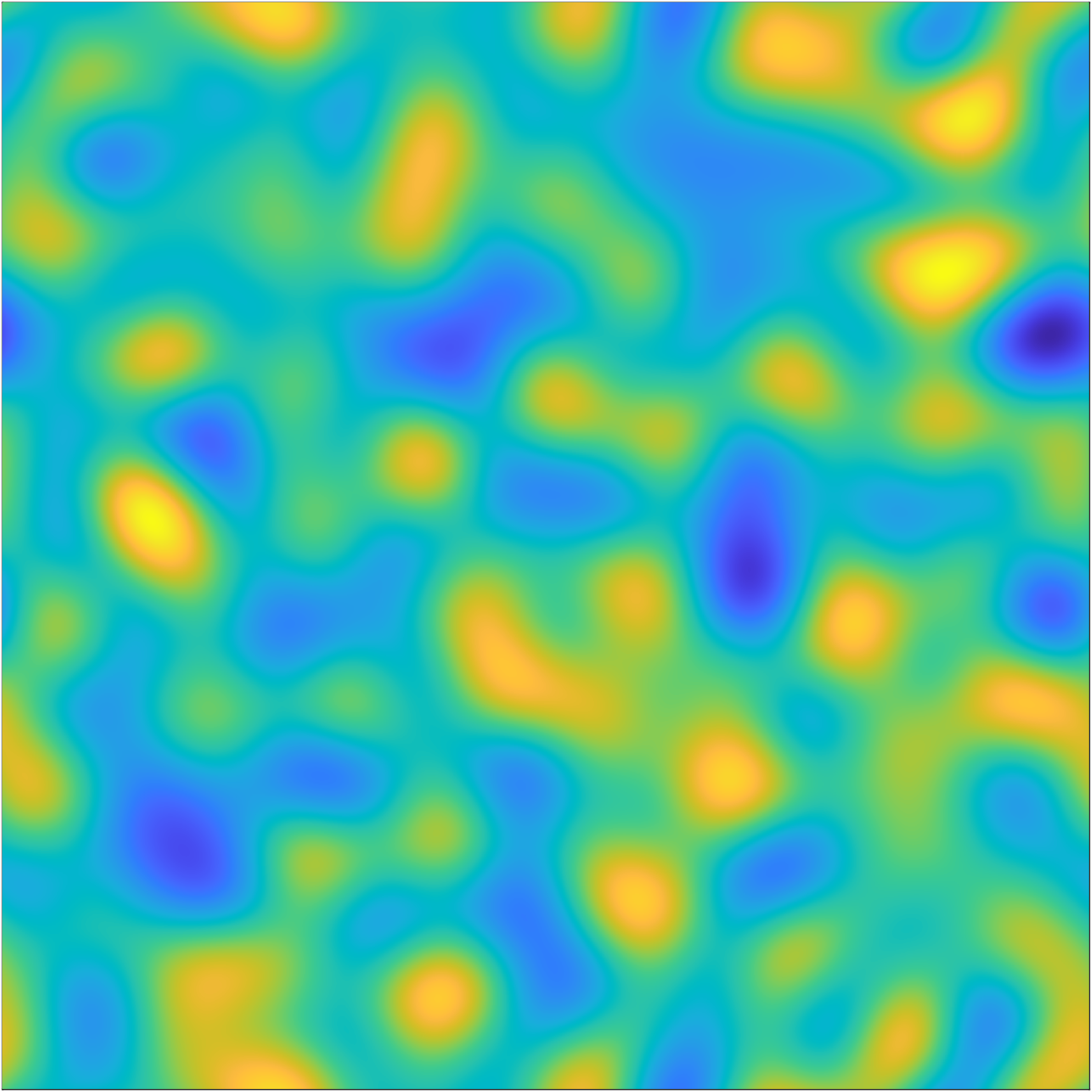}
    	\includegraphics[width=0.5\linewidth]{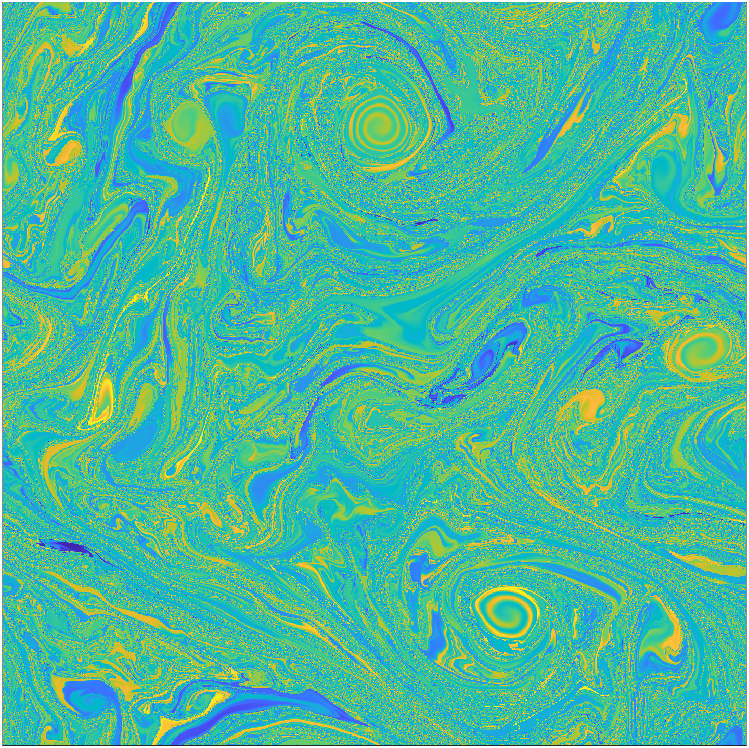}
     }
    \subfigure[]{
    	\includegraphics[width=0.5\linewidth]{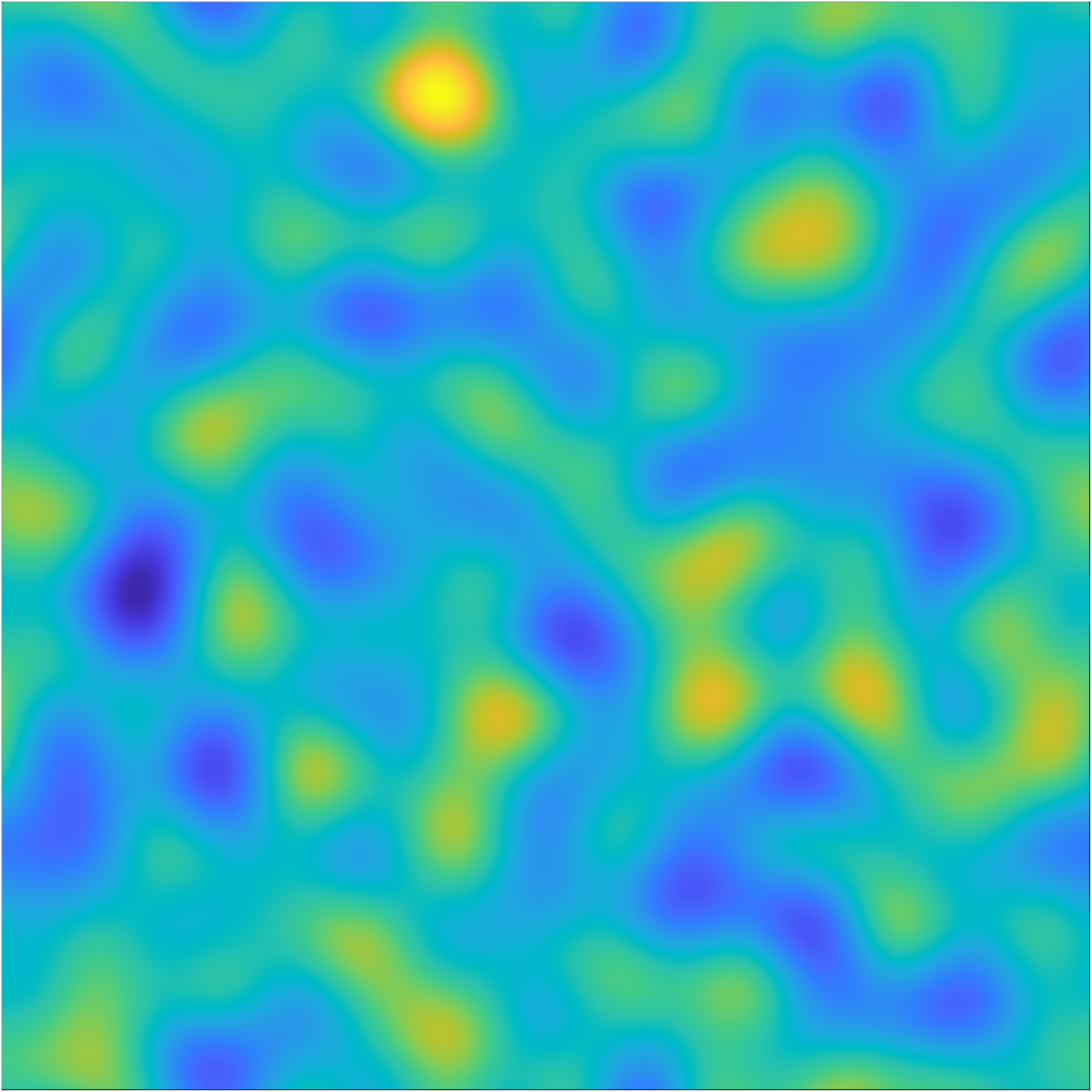}
    	\includegraphics[width=0.5\linewidth]{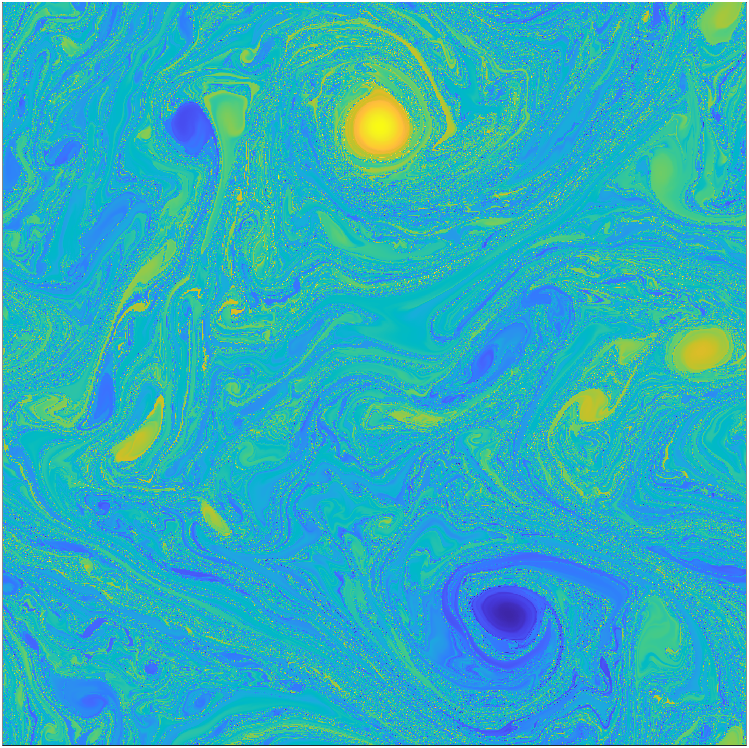}    	
    }
    \end{minipage}

\caption{\small Results from CMM computations. Figure (a) shows the final time scalar spectra for various correlation values. Figure (b) shows the mixing enhancement $M(\phi, t_*)$ as function of the scalar-vorticity correlation when transported by various flows for a given amount of time. The average enhancement over a sample of 30 scalars is shown in full line, the shaded area covers one standard deviation. Figures (c) and (d) show the initial and final time scalar fields for the $h=0$ and $h=1$ cases respectively.}
  \label{fig:CMM}
\end{figure*}

What we can retain from the results of equilibrium statistical mechanics is that it indicates a tendency of less efficient mixing when the vorticity-scalar correlation is strong. The worst mixing is attained when the velocity scale is largest, $k_u/k_0\approx 1$. We note that this is a relevant case, since two-dimensional flows are often dominated by space-filling structures.

We will assess now if, as suggested by statistical mechanics, mixing is reduced when $h\neq 0$, dynamically, when the system is not Galerkin truncated. Simulating the evolution of vorticity and the mixing of a scalar without the influence of viscosity and diffusion is numerically challenging. Nevertheless, recent advances in numerical analysis \cite{mercier2020characteristic} have led to the development of a numerical method allowing to compute, in a precise and efficient way, exactly this evolution. The method we therefore apply is the characteristic mapping method \cite{yin2021characteristic}, a method which, instead of advancing the Eulerian fields such as vorticity and velocity, evolves the inverse flow-map.
An advantage of this method is that once the flow map 
is determined for a given flow, it allows to determine directly the scalar field $\phi(\bm x,t)$ for all possible initial conditions $\phi_0(\bm x)$. Furthermore, the solution of this advection is not frequency truncated and solves the non-diffusive advection. The Casimirs and invariants of the system are naturally conserved so that we can compare the statistical predictions for the truncated system with the long-time solutions of the non-truncated one.

We start from an initial velocity field given by an enstrophy spectrum concentrated around $k_u$ with random phases and normalized to unit kinetic energy. The initial scalar spectrum is defined by the bump function with $k_\phi \approx 3.3$. The evolution of the energy spectrum and the scalar spectrum are shown in Fig.~\ref{fig:CMM}(a) for a range of correlation values. We see that as correlation decreases, more scalar energy is moved towards the small scales indicating a better mixing.

We quantify the extent of the mixing enhancement using the mixing norm of $\phi$. Define the enhancement
\begin{equation}
M(\phi, t) = \frac{k_{\phi(t)}}{k_{\phi(0)}} = \sqrt{\frac{\int | \nabla^{-1} \phi(0)|^2 d \vx }{\int | \nabla^{-1} \phi(t)|^2 d \vx }} ,
\end{equation}
where the second equality is obtained from the fact that $\int \phi^2 d \vx$ is a conserved quantity. The enhancement number $M(\phi, t)$ measures the relative increase of the typical wavenumber in time due to transport by the fluid. The essential difference with the results of equilibrium statistical mechanics is that the enhancement measure $M(\phi, t)$ takes into account the initial conditions, whereas equilibrium statistical mechanics only gives information on the final state, irrespective of initial conditions.
We note that in the $h=1$ fully correlated case, $\phi = (Q/W) \omega$, in which case $k_\phi = k_u = \sqrt{W/E}$ which is a constant of motion and hence $M(\phi, t)$ must remain 1. 

To test the influence of the scalar-vorticity correlation on the mixing enhancement, we perform the scalar advection using three flows obtained from the Euler equations where the characteristic length scale $k_u$ is approximately $1.7, 3.3$ and $9.4$ respectively. We generate for each flow a sample of 30 initial conditions $\phi'_0$ of random phase and bump function spectrum such that $\phi'_0$ are fully decorrelated from the vorticity $\omega_0$ and $k_{\phi'_0} \approx 3.3$. Each $\phi'$ correspond to the $h=0$ case, and together with $\omega$, we generate a family of initial conditions $\phi$ ranging from $h=0$ to $h=1$ and whose initial characteristic length $k_{\phi_0}$ range from $k_{\phi'_0}$ to $k_u$. This effectively tests for the mixing enhancement generated by a flow which moves at larger, same, and smaller spatial scales than the advected quantity. Fig.~\ref{fig:CMM} (b) shows the results from our tests. The time $t$ at which the enhancement is assessed is fixed at $t \approx 20 t_*$ with $t_* = W^{-1/2}$ as to normalize the average turnover over the three flows. The same tests can be performed by instead scaling $t$ with $E^{-1/2}$ but have similar results due to the scaling symmetry of the Euler equations. In all cases, we observe that the mixing enhancement is monotonously decreasing with respect to the correlation.

\paragraph{Conclusion.}

The fact that initial velocity conditions are important for the efficiency of  mixing is not surprising. However, the fact that for a given random scalar field and a given random velocity field, fixing $E,W,S$, the mixing can as dramatically be affected by the value of $Q$ as observed in Fig.~\ref{fig:CMM} is something which might not have been expected from the outset. Indeed, the mixing scale for $h=1$ remains constant, while for the fully uncorrelated case the mixing enhancement, for all our tests, is typically a factor 2. 
The underlying reason, which we now understand, is that, if the vorticity is fully correlated with the scalar, the scalar cascade is entirely locked to the enstrophy cascade. It seems from the test cases considered here, that due to its link with velocity field, the vorticity is indeed the least mixed scalar field in 2 dimensional flows. 

In the definition of the mixing-scale and the invariants of the system, we have not explicitly assumed that the system is turbulent. We do therefore think that these ideas can be applied to non-turbulent systems as well.  A practical implication of the present insights is that, in optimization problems (e.g. \cite{lunasin2012optimal,lin2011optimal,foures2014optimal}), the constraint of minimizing the correlation between scalar and vorticity, might be a powerful way to approach the optimal mixing regime in both turbulent and non-turbulent flows.

{\it Acknowledgments}. All DNS simulations were carried out using the facilities of the PMCS2I (\'Ecole Centrale de Lyon). For the purpose of Open Access, a CC-BY public copyright licence has been applied by the authors to the present document and will be applied to all subsequent versions up to the Author Accepted Manuscript arising from this submission.

\bibliographystyle{unsrt}
\bibliography{VorticityScalarCorrelation}

\end{document}